# Investigating the Potential of Test-Driven Development for Spreadsheet Engineering

Alan Rust, Brian Bishop, Kevin McDaid
Dundalk Institute of Technology,
Dundalk, Ireland
alan.rust@dkit.ie, brian.bishop@dkit.ie, kevin.mcdaid@dkit.ie

**ABSTRACT**

*It is widely documented that the absence of a structured approach to spreadsheet engineering is a key factor in the high level of spreadsheet errors. In this paper we propose and investigate the application of Test-Driven Development to the creation of spreadsheets. Test-Driven Development is an emerging development technique in software engineering that has been shown to result in better quality software code. It has also been shown that this code requires less testing and is easier to maintain. Through a pair of case studies we demonstrate that Test-Driven Development can be applied to the development of spreadsheets. We present the detail of these studies preceded by a clear explanation of the technique and its application to spreadsheet engineering. A supporting tool under development by the authors is also documented along with proposed research to determine the effectiveness of the methodology and the associated tool.*

## 1. INTRODUCTION

End-user programming has become the most common form of programming in use today [Burnett, 2004] with commercial-off-the-shelf spreadsheet packages one of the most popular program categories for end-user programming. The ubiquity of spreadsheet programs within all levels of management in the business world indicates that important decisions are likely to be made based on the results of these, mainly end-user developed, programs. The financial sector is particularly dependent on spreadsheets [Croll, 2005]. Unfortunately, the quality and reliability of spreadsheets is known to be poor following empirical and anecdotal evidence collected on the subject [Panko, 1998].

We propose the application of the software development methodology, Test-Driven Development (TDD) [Beck, 2003], [Ambler, 2006] to spreadsheet engineering. TDD is a proven technique for the development, rather than testing, of software that originated as a best-practice in Extreme Programming [Beck, 2000]. Extreme Programming is one of a number of software development processes, known as Agile Methods, that advocate a more flexible customer focused approach to the creation of software.

TDD is a coding technique that insists that the software developer writes the tests before they write the code. Code is then written to pass these tests keeping the code as simple as possible. It is rewritten, known as refactoring, if the design becomes unwieldy. TDD, supported by a dedicated tool, has been shown in software engineering to improve the quality of code and to support the testing and maintenance of software. In this paper we show that the technique can be applied to the development of spreadsheets.

The layout of the paper is as follows. Section 2 introduces the topic of spreadsheet engineering with a focus on the current poor quality of spreadsheets and the attempts to apply software engineering practices to their development. Section 3 explains the Test-Driven Development methodology and includes a detailed explanation of how TDD can be applied to the development of spreadsheets. Tools that support TDD are discussed in Section 4, as are details of a tool developed by the authors to support TDD for spreadsheets. In Section 5, the results of two case studies in which spreadsheets were developed using the TDD tool and methodology are described. The lessons learned from these case studies are presented and discussed. Section 6 concludes the work with a discussion of the potential of TDD in the context of related work. Prospective future research is also outlined.

**2. SPREADSHEET ENGINEERING**

It is estimated that by 2012 there will be 90 million end-users in U.S. workplaces, of which 55 million will use spreadsheets and databases, with potentially 25 million performing programming-like spreadsheet and database activities. The number of end-user developers will vastly exceed that of professional software developers who are estimated to number less than 3 million by the same year [Scaffidi, 2005].

End-user developers use little or no formal software development processes and adhere to almost none of the software industry standard development practices. Although there is little evidence to suggest that cost overruns or schedule irregularities in end-user development (EUD) are a major concern, it is widely accepted and documented that the reliability of software resulting from EUD, especially spreadsheets, is dangerously low. Numerous works have addressed this issue, [Panko, 1998], [Rajalingham, 2000] and [Chadwick, 2004], and have concluded that spreadsheets have a notoriously high number of faults, and very little testing, formal or otherwise, is ever carried out.

One particular study of spreadsheet errors based on the experience of a consulting firm in England, Coopers and Lybrand, concluded that 90% of spreadsheets with over 150 rows of data were found to contain one or more faults [Panko, 1998]. When spreadsheet failures do occur, the results can be quite significant. For example, sudden budget cuts were necessary at the University of Toledo after an erroneous spreadsheet formula inflated projected annual revenue by $2.4 million [Fisher, 2006]. Another spreadsheet error caused stock values at Shurgard Inc to fall sharply after two employees were overpaid by $700,000, and a cut-and-paste spreadsheet error cost the Transalta Corporation $24 million through overbidding [Fisher 2006].

It has been proposed that poor spreadsheet reliability is due to the absence of structured methodologies in the design, implementation and testing of spreadsheets and the general lack of formal policies in organizations for the development of spreadsheets [Panko, 1998], [Rajalingham, 2000]. This is regardless of the fact that the organizations may rely on spreadsheets for critical day to day business activities. Apart from the problems caused by lack of end-user development methodologies, the inherent flexibility of the spreadsheeting environment may also contribute to the problem.

Spreadsheet engineering is concerned with all aspects of creating spreadsheets and in practice is the application of traditional software engineering principles to spreadsheet development [Grossman, 2002]. Numerous methodologies and principles have been proposed for spreadsheet development. These range from variations of traditional software engineering methodologies [Rajalingham, 2000] such as the waterfall model, to the application of the principles of Extreme Programming (XP) to spreadsheet development [O'Beirne, 2002]. While each separate body of research may present or propose a different approach, they generally agree that the absence of a structured approach to spreadsheet engineering is one of the main reasons behind spreadsheet errors.

Spreadsheets have been referred to as the original agile development environment [O'Beirne, 2002], and it has been argued that agile methodologies may be better suited to end-user development than more traditional methodologies [Segal, 2004], [Grossman, 2002]. Agile Methods involve minimal documentation and, to be successful, are based on small development teams liaising closely with the customer. They are particularly popular for software projects of short duration.

Extreme Programming, the most popular Agile Method, recommends a number of best practices. One such practice is pair programming where developers work together when coding the system [Beck, 2000]. The application of a similar practice, termed Cooperative Debugging, to spreadsheet development has been explored in [Panko, 1998]. We propose the application of a second Extreme Programming best practice, namely Test-Driven Development (TDD), to the creation of spreadsheets. Note that although Test-Driven Development is one of the most successful agile practices it can also be used within a plan driven methodology.

### 3. TEST-DRIVEN DEVELOPMENT

In recent years Test-Driven Development has emerged as one of the most successful developer enhancing productivity techniques. The easiest way to describe Test-Driven Development is as follows – write a failing test, make the test run, then finally refactor the code so that the design becomes cleaner. This process is then repeated over and over until the system is completed.

Numerous studies have been carried out to compare Test-Driven Development with more traditional development techniques such as code inspections or testing after code has been written. One such study showed that Test- Driven Developers produced a "higher quality of code, which passed 18% more functional black box test cases and 80% of the developers who took part in the survey believed Test-Driven Development was more effective" [George, 2003].

Test-Driven Development works because it forces the developer to consider the design, through the tests, of the system element before coding and to take small steps when writing the software. It also stresses simple solutions to problems.There are ancillary benefits to TDD. Not only has it been shown that Test-Driven Development improves the quality of code, the tests themselves can also be used as system level documentation [Ambler, 2006]. Also, as the tests are automated, the system is easy to retest following the addition of new functionality. Manually running tests would of course increase project time substantially.

Dedicated tool support has been central to the success of TDD. XUnit is the collective name for the various tools that have been created to support Test-Driven Development. There are tools to aid programming in Java (JUnit), VB (VBUnit) and VBA (VBAUnit) to name but a few. These tools allow the developer to create a complete test suite which can be run at any time. The tools use a very simple method to show if the tests pass or fail. A green light is used if tests pass and a red light is shown if tests fail. It is important to remember that new functionality will only be added if all the tests run successfully.

### 3.1 Applying Test-Driven Development to Spreadsheet Engineering

Figure 1 [Ambler, 2006] shows how a Test-Driven Development project should proceed. The first step in Figure 1 is to add a test which tests some sort of functionality needed in the system. The test is then run; this test will of course fail as the code to pass the test has not yet been created. The next step is to create the code which will implement the required functionality i.e. write the code which will make the test pass. After the code has been created the test is run again. If the test passes the developer can go on to add new functionality or they can choose to refactor the code written so that it becomes cleaner. However if the test fails the developer must go back to the code and make the changes required to pass the test. After the changes have been made the test is run again. Again if the test passes, development can continue and if the test fails more changes must be made to the code. This process is then repeated until all the required functionality of the system has been implemented.

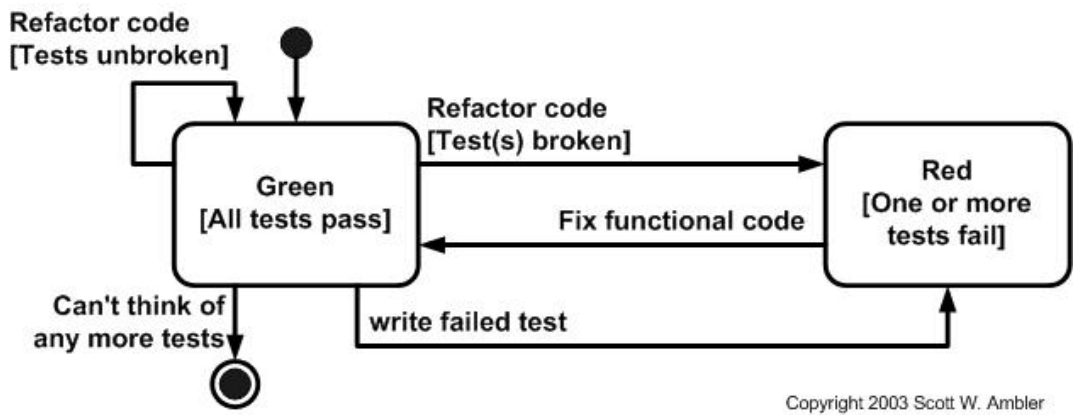

**Figure 1 - Test-Driven Development**

The following example shows how TDD can be applied to Spreadsheet Development.

*Example*
Assume that we wish to input a suitable formula in a cell B2 to convert a student mark in cell A2 to a grade according to the following rule:
- A mark below 40 yields a "FAIL" grade.
- A mark of 40 or more up to but not including 70 yields a "PASS" grade.
- A mark of 70 or more up to and including 100 inclusive yields a "HONOR" grade.

Using the Test-Driven method discussed above we create our initial tests. Looking at the problem it is clear that to start with we need to create 3 tests. These tests will cover the 3 grade levels. We set the three tests up, using a tool we shall describe later, with input mark values of 20.5, 55.31 and 78.85 which should result in output grades of Fail, Pass and Honor respectively. Again using the tool we run these tests which will of course fail as no formula has been added to cell B2. Note that any value can be placed in cell A2. To make sure that the tests are working correctly we place "FAIL", "PASS" and "HONOR" in turn into the cell B2 and re-run the tests. The tests should now in turn pass and we therefore know that they are working correctly.

The key is to build the formula up in a series of short steps. It is best to start with a formula that will pass the first test i.e. the Fail test. The formula may be

=IF(A2<40,"FAIL","PASS")

We run the tests again and note that the first and second tests now pass but the third test is still failing. We therefore need to go back to the code and make some changes to the formula. The new form may be as in Cell B2 of Figure 2.

|   | A | B |
|---|---|---|
| 1 | Mark | Grade |
| 2 | 0 | =IF(A2<40,"FAIL",IF(A2<70,"PASS","HONOR")) |

**Figure 2 – Formula for Grade that passes first 3 tests**

Again the tests are run with all three passing. However we are well aware that we have only tested for values that are within the required ranges. We should also test for values that lie on the boundary of our ranges. Here we can create a number of new tests which should include the following:

- A test with an input value of 0 with an expected output value of "FAIL".
- A test with an input value of 40 with an expected output value of "PASS".
- A test with an input value of 70 with an expected output value of "HONOR".
- A test with an input value of 100 with an expected output value of "HONOR".

All five tests are run with both the original and additional tests passing. We are now happy with the basic calculation of the grade but are aware that the formula should deal appropriately with inputs that are not in the range 0 to 100. Assuming we want "Not Valid" to appear if the user enters an invalid entry, we create more tests as follows to add this extra functionality.

- A test with an input value of -1.5 with an expected output value of "NOT VALID".
- A test with an input value of 124.45 with an expected output value of "NOT VALID".
- A test with an input value of "TEST" with an expected output value of "NOT VALID".

Again all the tests are run. Our original tests still work correctly but our 3 new tests fail. Therefore development can only continue when these new tests pass. So we have to modify or refactor the formula to something like this:

=IF(AND(A2<40,A2>=0),"FAIL",IF(A2<70,"PASS","HONOR"))

Again run all tests, we see that all tests are passing apart from the new ones. So we go back to the code again and make more changes.

=IF(AND(A2<40,A2>=0),"FAIL",IF(AND(A2>=40,A2<70),"PASS","HONOR"))

We get the same result this time as we did after our last change but we know we are closer to a completed formula. The last change is made with the resulting formula shown in Figure 3:

|   | A | B |
|---|---|---|
| 1 | Mark | Grade |
| 2 | 0 | =IF(AND(A2<40,A2>=0),"FAIL",IF(AND(A2>=40,A2<70),"PASS",IF(AND(A2>=70,A2<=100),"HONOR","NOT VALID"))) |

**Figure 3 – Final formula for determination of grade**

Finally we run the tests and we see that all ten pass. We have implemented our original specification and can now continue to add extra functionality to the system in the same way.

When initially presented the incremental nature of TDD often appears inefficient to developers. However, with practice, developers learn to build the tests and code very quickly. As they become more experienced developers can, if they are very sure of the solution, use fewer steps in the process. However, the insistence on coding through small steps with regular checks on progress guards against the temptation, buoyed by overconfidence, to implement large numbers of formulas without intermediate checks. In this way TDD has the potential to address the clear issue of developer overconfidence in the creation of spreadsheets, a topic researched by a number of authors including [Thorne, 2004].

**4. TOOL SUPPORT FOR TEST-DRIVEN DEVELOPMENT**

The importance of tool support for Test-Driven Development cannot be overstated. As already mentioned manually running tests would increase project time substantially. Therefore the so called XUnit tools such as JUnit provide an automated way to run all the created tests. To show the correctness of tests a simple system is used. A green light is shown for correct tests and a red light is shown for tests that are failing. The tests themselves can be run individually or they can be created within a test suite and run all at once. JUnit has the following features:
- Assertions for testing expected results
- Test fixtures for sharing common test data
- Test runners for running tests

The tests in JUnit are written in Java code and similarly the tests in VBUnit are written through VB Code. This is the main difference when our tool is compared to other XUnit tools. Our tests are generated automatically in the background. This means that the developer does not need to worry about how the tests are created, they just need to think about what tests need to be created. Figure 4 shows a screenshot of the TDD tool developed for use with Microsoft Excel and currently undergoing improvements.

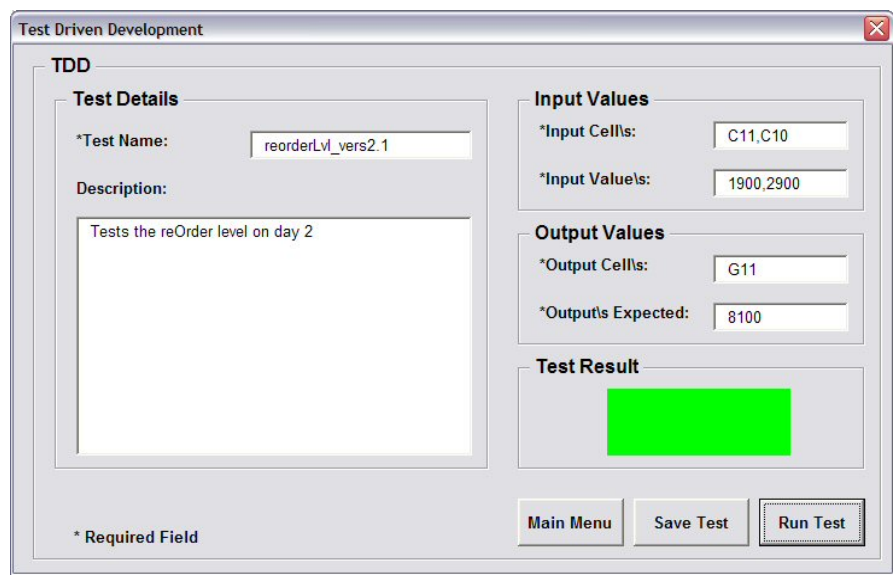

**Figure 4 – Screenshot of TDD tool**

Figure 4 shows a screenshot of the tool with details of a single test named "reordLvl_vers2.1". For this test the output cell, which will contain the formula to be developed, is G11 with inputs coming from cells C11 and C10. The test is whether G11 gives a value of 8100 when C11 and C10 have the values 1900 and 2900 respectively. When the test is run the tool substitutes 1900 into Cell C11 and 2900 into cell C10. The value in G11, based on the formula present, is then checked to make sure it is equal to 8100. In this case the value was correct and the test result is a green light. If the value were incorrect a red light would be displayed and further changes would be made to the code until the test passes. After the test is run, the original values in the input cells are re-entered. Future work on the tool is outlined in Section 6 following discussion of the case studies. Currently the tool allows both text and number entry and can take inputs from multiple sheets.

**5. RESEARCH QUESTION**

On the basis of its success in software engineering the authors propose the use of Test-Driven Development in the creation of spreadsheets. A tool, whose initial design as explained in Section 4 mimics the established TDD tools, has been created to support the methodology. While an example of how TDD can be applied was presented in Section 3 the question remains as to whether TDD has the capacity to reduce the level of spreadsheet errors.

The issue as to which types of error may be reduced is also of interest. Our belief is that TDD will naturally target mechanical and logic errors as described in [Panko 1998]. Potentially, through the process of specifying tests before coding it also may be able to reduce omission errors. Furthermore, input value errors could also be checked and thus possibly reduced through tests that examine single input cells.

Even if the methodology is shown to reduce errors there are a myriad of human factors that must be addressed before it is likely to be adopted by practitioners. The question as to whether it is more suited to End-User development as opposed to product/bespoke activity and whether it is more beneficial to certain types of spreadsheet are also important. A key question may be the effectiveness of the methodology in terms of the trade-off between the (hoped for) improved reliability and the likely additional cost (in terms of time) of applying the approach.

We propose to answer these questions through a series of structured trials involving both students and industry spreadsheet developers. Before structured experiments designed to answer this question can be conducted, it is important to establish that TDD can be applied in the spreadsheet domain and to ascertain if there are any particular questions or considerations that need to be addressed. Issues regarding the tool support required for Spreadsheet TDD are also important. An initial investigation comprised of two case studies was conducted to identify these issues.

In the first case study the first author, a novice spreadsheet developer with a background in software programming, used TDD to create a small inventory spreadsheet to model stock levels. The third author, an experienced spreadsheet developer with no formal training in programming, used TDD in the larger second case study to develop a spreadsheet to track progress during software projects that utilise an agile development process. It is intended that this spreadsheet will be made available to companies through the SPACE project [SPACE, 2006] that seeks to promote agile methods in the software industry.

**5.1 Case Study 1**

The first case study was based on a hospital inventory system used to predict the likelihood of stock levels falling below a critical value. Complicating factors involved the time to deliver stock and the batch size for orders. The small single sheet system was developed by the first author in Microsoft Excel using the plug-in tool described.

By the time the small system was finished there were a total of 15 tests. They test the entire system from opening stock levels on day1 to the percentage of days that fall below the critical point. The developer, a relative spreadsheet novice with little practical experience of TDD, began by identifying the main inputs that would be needed in the system and built tests around these. As more system functionality was addressed further tests were added. Many of the tests included a large number of inputs. Overall, the developer found that this method really helped in creating the code. Instead of having a specification in his head of how to code the problem, the specification for the function was in the form of tests. Therefore after each new function was added and all the tests passed the developer felt confident that he could move on and add in the next part.

While the developer felt that the approach meant that he was more confident of the spreadsheet's correctness there were a number of issues. The first related to the difficulty of testing random numbers generated to simulate stock changes. The second and third relate to the tool and the need to improve it to allow for improved entry of values when the number of input and output cells are high and to automate the copying of tests in the case where output cells are to be copied. We now turn our attention to the much larger Case Study 2.

**5.2 Case Study 2**

In Case Study 2 the third author developed a spreadsheet in Microsoft Excel to track the progress of software projects utilising an Extreme Programming process with weekly iterations. The spreadsheet is adapted from a couple of such spreadsheets currently in use by software firms. The developed spreadsheet contained two sheets, one with the project and tracking details and the other with the calculation and graphical display of progress. A print-friendly screenshot of the first sheet is shown in Figure 5 below. The shaded cells represent the numeric/text values required for input by the project manager.

The system contained 246 cells containing formulas. Of this number there were 18 distinct formula cells with the remainder of cells filled by copying operations. In the development of the spreadsheet a clear issue arose with the tool and the need to include functionality to copy and paste tests specified for one cell to another cell thereby mimicking the functionality of the spreadsheet. Of the 18 distinct formula cells, some of which were quite involved, the number of tests written for each ranged from 1 to 8. These tests changed considerably as the design of the formulas in the cells matured. Were the tests to be repeated for each of the copied cells there would have been in excess of 500 tests in total.

As with Case Study 1 the developer felt that the methodology and tool worked well and that there was an increased confidence in the reliability of the spreadsheet following the adoption of the approach. In fact the developer felt that were he to start again he would write an even higher number of tests for some of the key formulas. Interestingly he also felt that he would refactor at least two of the formulas to write them as macro functions.

The case studies revealed a number of key issues. Amongst those relating to the tool are the following:
- A feature is required of the tool to support the copying of tests to reflect the key copying and fill features of spreadsheets.
- The specification of a large number of inputs is difficult. At present the tool only allows for the manual entry of cells and cell values. Once a range has been entered it would be useful to have the option to use the original values in the cells rather than entering them manually. An interface similar to the what-if scenario feature in Microsoft Excel may be appealing.
- It is important to be able to group tests together and to run them in batches to test each function individually. This functionality should be added.
- It would be nice if the tool indicated which cells were the subject of tests and which were not. Functionality which showed, through red and green colour coding, whether these were passing tests or not associated with formula cells would be ideal.

Other issues pertaining to the methodology include:
- If possible the need to extend TDD to address incorrect input and omission errors as well as logical errors. It may be possible to reduce input errors by writing tests that specify the values that should be in individual cells.
- How can TDD be applied to the development of graphical outputs such as time series graphs?

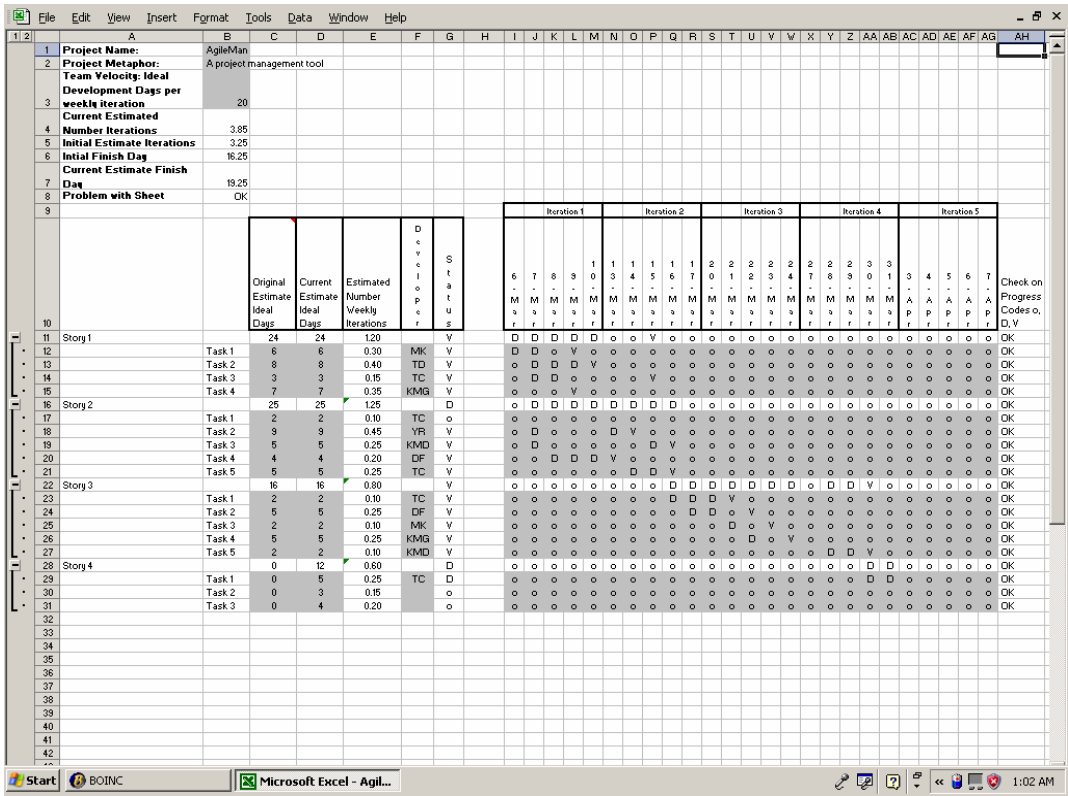

Figure 5: Agile Tracking Tool

## 6. CONCLUSION

This paper explores the potential of Test-Driven Development (TDD), a best-practice in Extreme Programming, to improve the engineering of spreadsheets. Through two case studies the authors have increased their understanding of TDD and how it can be applied in the spreadsheet domain. *Importantly, the authors have concluded that the methodology has the potential to improve the development of spreadsheets*. However, the studies have revealed a number of issues with the methodology but more so with the Microsoft Excel plug-in tool developed to support the TDD approach. These issues are currently being addressed before trials involving real users can be conducted to prove the effectiveness and efficiency of the innovative method.

While Test-Driven Development is a development rather than a testing approach, in practice the resulting automated test suite acts as a test harness for the system. An interesting question relates to the additional tests required beyond the TDD set. Answers to this question may well be based on such works as [Fisher, 2006] and [Pryor, 2004]. Similarly, while it can not replace documentation, a secondary benefit of TDD in software engineering is its ability to shine a light on the design and restrictions of the system. The question as to whether this can be the case for spreadsheet engineering is an open one as is the question as to whether TDD can aid the maintenance of spreadsheets.

## 7. REFERENCES

Ambler, S. W. (2006), "Introduction to Test Driven Development (TDD)", http://www.agiledata.org/essays/tdd.html, 30/03/2006 15.30

Beck, K. (2000), "Extreme Programming Explained: Embrace Change", Addison-Wesley, Pearson Education

Beck, K (2003), "Test Driven Development: By Example", Addison-Wesley

Burnett, M., Cook, C., Rothermel, G. (2004) "End User Software Engineering", Communications of the ACM, September, Vol.47, No. 9

Chadwick, D. (2004), "Stop That Subversive Spreadsheet", EuSpRIG Conference 2004, Program, Abstracts and Outlines

Croll, G. (2005), "The importance and criticality of spreadsheets in the city of London". Proceedings of the European Spreadsheet Risks Interest Group, 2005.

Fisher, M., II, Rothermel, G., et al, (2006), "Integrating Automated Test Generation into the WYSIWYT Spreadsheet Testing Methodology", ACM Transactions on Software Engineering and Methodology (to appear), http://web.engr.oregonstate.edu/~burnett/reprints.html 30/03/2006 15.30

George, B. and Williams, L., (2003) "An Initial Investigation of Test-Driven Development in Industry", Proceedings ACM Symposium on Applied Computing, Melbourne, FL

Grossman, T. A. (2002), "Spreadsheet Engineering: A Research Framework" European Spreadsheet Risks Interest Group 3rd Annual Symposium, Cardiff, pp. 21-34, July

Musa, J. (1999), "Software Reliability Engineering", McGraw-Hill Publishing

O'Beirne, P. (2002), "Agile Spreadsheet Development (ASD)", Systems Modeling Ltd., http://www.sysmod.com/agile.htm, 30/03/2006 15.30

Panko, R. (1998), "What We Know About Spreadsheet Errors", Journal of End User Computing 10, 2 (Spring), p15–21

Pryor, L. (2004), "When, Why and How to Test Spreadsheets". Proceedings of the European Spreadsheet Risks Interest Group Symposium, 2004.

Rajalingham, K., Chadwick, D., Knight, B., et al. (2000), "Quality Control in Spreadsheets: A Software Engineering-Based Approach to Spreadsheet Development", Proceeding of the 33$^{rd}$ Hawaii International Conference on System Sciences, IEEE. Vol. 33

Rajalingham, K., Chadwick, D., Knight, B., et al. (2000), "Classification of Spreadsheet Errors", European Spreadsheet Risks Interest Group Symposium, Greenwich, England

Scaffidi, C., Shaw, M. and Myers, B. (2005), "The "55M End-User Programmers" Estimate Revisited", Institute for Software Research International Technical Report CMU-ISRI-05-100

Segal, J. (2004), "Professional end user developers and software development knowledge", Technical Report 2004/25, Department of Computing, The Open University, UK

SPACE Project, (2006), http://www.agileireland.com/, 30/03/2006 16.34

Thorne, S., Ball, D., and Lawson, Z. (2004), "A novel approach to formulae production and overconfidence measurement to reduce risk in spreadsheet modeling", European Spreadsheet Risks Interest Group Symposium.

**Blank page**